# How circular economy and industrial ecology concepts are intertwined? A bibliometric and text mining analysis

Michael Saidani[1]*, Bernard Yannou[2], Yann Leroy[2], Francois Cluzel[2], Harrison Kim[1]

[1]*University of Illinois at Urbana-Champaign, Industrial and Enterprise Systems Engineering, USA*
[2]*Université Paris-Saclay, CentraleSupélec, Laboratoire Genie Industriel, Gif-sur-Yvette, France*
*Contact author: msaidani@illinois.edu*

**Abstract**

Combining new insights from both bibliometric and text mining analyses, with prior relevant research conversations on circular economy (CE) and industrial ecology (IE), this paper aims to clarify the recent development trends and relations between these concepts, including their representations and applications. On this basis, discussions are made and recommendations provided on how CE and IE approaches, tools, and indicators can complement each other to enable and catalyze a more circular and sustainable development, by supporting sustainable policy-making and monitoring sound CE strategies in industrial practices.

## 1. Introduction

While industrial ecology (IE) has been thoroughly theorized and analyzed over the last two decades (Cecchin et al., 2020), the research around the circular economy (CE) is still emerging and gains increasing traction among businesses, policymakers, and academics (Korhonen et al., 2018a). Yet, this CE concept is not completely new and finds its foundations in several research streams, including the IE. For instance, according to the Ellen MacArthur Foundation (EMF, 2015), the CE paradigm is based on seven "schools of thought", namely: industrial ecology, biomimicry, natural capitalism, regenerative design, cradle-to-cradle, blue economy, performance economy. For the French Environment and Energy Management Agency (ADEME, 2014), industrial ecology is also one of the several pillars to build a circular economy, among ecodesign or product-as-a-service.

In the meantime, it appears that IE has found a second wind in the light of the actual momentum around the shift towards a more CE (Cecchin et al., 2020). Moreover, this inclusion of IE within the CE framework is not acknowledged by all. In fact, the connections between CE and IE are still stirring a vivid debate between scholars, through opposite scientific papers (Buclet, 2015; Figuière and Chebbi, 2016), or by using the online exchange platform ResearchGate (2019) to figure out, e.g., "are circular economy and industrial ecology the same concept?". Also, a special issue of the Journal of Industrial Ecology entitled "Exploring the Circular Economy" published in 2017 (Bocken et al., 2017) brought the question of whether industrial ecology is the science of the circular economy.

As circular economy and industrial ecology are therefore still two moving and fuzzy concepts, with no clear boundaries or standardized definitions, this paper aims to brings new insights to the following questions: (i) is circular economy the new industrial ecology, or to what extent it can be considered as such?; and (ii), importantly, how these concepts can be exploited or combined together to achieve a more sustainable development?

In line with the statement of Lifset and Graedel (2002) at the beginning of the 21st century and emergence of industrial ecology – *"Setting out the goals and boundaries of an emerging field is a hapless task. Set them too conservatively and the potential of the field is thwarted. Set them too expansively and the field loses its distinctive identity. Spend too much time on this task and scarce resources may be diverted from making concrete progress in the field. But in a field with a name as provocative and oxymoronic as industrial ecology, the description of the goals and definitions is crucial"* – the present piece of research aims to provide an update on the positioning between the industrial ecology and circular economy concepts, based on both the pioneered and latest publications covering these topics, using bibliometric and text mining tools to ensure an objective standpoint.

## 2. Methodology

First, prior relevant works dealing with the analysis of both circular economy (CE) and industrial ecology (IE) concepts are reviewed and summarized. Then, after identifying and extracting the diverse definitions of a CE and IE, a text mining tool is used to carry out a comparative semantic analysis of these definitions (i.e., to figure out what are the keywords that stand out and how are they related), in order to highlight possible convergences and differentiation. In parallel, a bibliometric analysis is conducted to depict and understand the evolution of published papers from 2000 to 2019 (in top-tier peer-reviewed journals), having the keywords "circular economy", "industrial ecology", or both. Finally, discussions are made on the contributions of these two fields to achieve sustainable development, e.g., to enable and exploit the full potential of ecodesign approaches.

Bibliometric analysis belongs to the scientometry research field, which addresses the quantitative analysis of activity and scientific networks (Leydesdorff and Van den Besselaar, 1997). Regarding the text mining analysis, the term frequency – being an efficient and straightforward text mining method (Gaikwad et al., 2014) – is the technique used here to analyze and compare the definitions of circular economy, industrial ecology.

Note that Deus et al. (2017) previously analyzed the current state of publications with the keyword "circular economy" (either located in the title, abstract, or keywords), using bibliometric tools, but without making any bridges with industrial ecology. Türkeli et al. (2018) also analyzed the evolution of the scientific knowledge on CE, produced in the European Union and in China (most productive regions in this field) using bibliometric, network and survey analysis. Interestingly, our updated and new findings considering both the evolution of publications in IE and CE research streams, are put in parallel with the insights generated by these previous but complementary bibliometric analyses.

## 3. Literature survey

### *3.1. Evolution: from industrial ecology to circular economy*

Although the initial ideas and first consistent mentions of industrial ecology can be traced back in the early 1950s, the official birth of the "industrial ecology" concept can be related to the scientific paper by Frosch and Gallopoulos (1989) that suggested the need for "an industrial ecosystem" in which "the use of energies and materials is optimized, wastes and pollution are minimized, and there is an economically viable role for every product of a manufacturing process", in accordance to the International Society of Industrial Ecology (ISIE, 2015). The overarching goal was to have industries working together in order to move from a linear to a cyclical closed-loop system. A concrete demonstration of this concept started in 1972 in Denmark, in Kalundborg, and is still ongoing.

Back in the 1990s, industrial ecology was considered in the research literature as an emerging framework. Erkman (1997) viewed industrial ecology as a means to implement sustainable considerations in an industrial society. To Røine (2000), the most critical challenge in industrial ecology was to unify two main interests: ecological sustainability on the macro level, and business economy profit on the micro level. This means that knowledge from different actors and disciplines are needed to implement necessary processes of change.

Meanwhile, Garner and Keoleian (1995) identified future needs for the development of industrial ecology, asking for a clearer definition of this field and its concepts, as it is the case at the moment for the circular economy. And as aforementioned in the introduction, Lifset and Graedel (2002) discussed the challenges of setting out the goals and boundaries of industrial ecology as an emerging field. As a new field, they considered that "industrial ecology is a cluster of concepts, tools, metaphors and exemplary applications and objectives". Note that during this period, several ecodesign methods and tools started to emerge to support the design and development of more sustainable products (Pigosso et al., 2010).

With this background, Ehrenfeld (2004) questioned whether industrial ecology was a new field or only a metaphor: "in the 10 years since industrial ecology first became a topic of academic interest, it has grown as a field of inquiry and has produced a community of practice in several sectors including academia, business, and government". To Ehrenfeld (2004), even as the shape of industrial ecology becomes clearer, ideas like industrial ecology must become institutionalized to have much effect on the reality of everyday activities.

In 2008, the circular economy paradigm had been being first institutionalized and materialized first in China, as part of the law entitled "Circular Economy Law of the People's Republic of China". The purpose of this law was to promote the CE to improve the use of resources and protect the environment and thus enable sustainable development. It defines the CE as "a generic term used to refer to all reduction, reuse and recycling activities carried out during the production, circulation and consumption process". On this basis, many countries started to incorporate some CE-related topics and objectives in their political agenda. For instance, in France, the CE concept was officially first mentioned during the "Grenelle de l'Environnement" and formalized in 2013 through the creation of the National Institute of Circular Economy (Bonet et al., 2014).

In a nutshell, Blomsma and Brennan (2017) depicted the stages the circular economy concept has gone through: the period from the 1960s to the 1980s is described as the preamble, the one from the 1990s to the 2010s as the excitement. According to the authors, today's period is facing the validity challenge, and forecasts three possible pathways for the CE: coherence, permanent issue, construct collapse.

### *3.2. Definitions of industrial ecology and circular economy*

In the 1990s, industrial ecology was theorized as "a new approach to the industrial design of products and processes and the implementation of sustainable manufacturing strategies" (Jelinski et al., 1992). It was defined as a concept in which an industrial system is viewed not in isolation from its surrounding systems but in concert with them – IE seeking to optimize the total materials cycle from virgin material to finished material, component, product, waste product, and to ultimate disposal. Accordingly, to Erkman (1997), IE is a study aimed at understanding the circulation of materials and energy flows; therefore, IE must first understand how the industrial ecosystem works, how it is regulated and its interactions with the biosphere in order to determine how the industrial ecosystem can be restructured to resemble how natural ecosystems function. Circular economy is also looking for an optimal management of resources throughout the life cycle of systems and is characterized by closed-loop systems, promoting maintenance, sharing, leasing, reuse, remanufacturing, and recycling. CE aims to retain the highest utility and value of products, materials, and resources at all times, in order to minimize the generation of waste (EMF, 2015).

Yet, as for now, there is no standardized or crystalized definition of the CE or IE, which might result in diverse interpretations of these concepts among different stakeholders, and therefore be a hindrance to their actual implementation in industrial practices. Based on their review of 25 definitions of CE, Sacchi Homrich et al. (2018) concluded that this concept came from different epistemological fields, and pointed out a lack of consensus and convergence in the literature. So far, the most comprehensive review has been performed by Kirchherr et al. (2017), identifying and scrutinizing 114 definitions of CE. An interesting finding is that 83 (73%) of these 114 definitions have been proposed between 2012 and 2017, showing how emerging and young the circular economy research field is. According to Kirchherr et al. (2017), such analyses are of the utmost importance to bring more coherence in the circular economy concept, because they argue that significantly varying CE definitions may eventually result in the collapse of the concept.

Blomsma and Brennan (2017) depicted the circular economy as an umbrella concept, which can be described as “a broad concept or idea used loosely to encompass and account for a set of diverse phenomena” (Hirsch and Levin, 1999). Indeed, applied to the circular economy, and the various resource strategies individually related to such a paradigm, this umbrella concept offers a new framing of these strategies by drawing attention to the relationship and complementary between these strategies. Furthermore, Kovács (2017) analyzed the CE concept in comparison with other related concepts that have been used in other disciplines, such as industrial ecology, or supply chain management, in order to understand what is novel, and how the circular economy extends or combines previous streams of literature. In this line, Blomsma (2018) provides an overview of waste and resource management frameworks, developed by a variety of actors, that either inspired the CE concept or that took inspiration from it.

In the following sub-section, the linkages, including convergence, divergence or complementary, between CE and IE are more precisely discussed based on a synthetic literature review. Then, further insights are provided in section 4, based on a new bibliometric analysis of the growing number of publications dealing with CE and IE, and a text mining analysis of their various definitions.

### *3.3. Connections between industrial ecology and circular economy*

To Bourg and Erkman (2003), it is not straightforward at first to see the differences between CE and IE, especially since, on the one hand, the main strategic axes of IE are defined as follows: (i) to value waste as a resource; (ii) to loop on material cycles and minimize dissipative emissions; (iii) to dematerialize products and economic activities; and, (iv) to decarbonize energy production. On the other hand, the French Environment and Energy Management Agency (ADEME, 2014) subdivides the CE into seven pillars, namely: (i) sustainable extraction, supply, procurement; (ii) ecodesign (of products and processes); (iii) industrial and territorial symbiosis; (iv) functional economy (i.e., product-as-a-service); (v) responsible consumption; (vi) longer duration of use; and, (vii) recycling and waste recovery. From this standpoint, IE is considered as one pillar of CE, but not that in this case, IE seems mainly to be reduced to the dimension of industrial symbiosis and eco-industrial parks. To Figuière and Chebbi (2016), a global approach is required to achieve the overall objectives of a CE, while a more local approach, e.g., on a region or industrial site, is more relevant for the realization of IE objectives. Also, from an ecodesign perspective, these pillars emerge naturally when considering the question of how to promote optimal use of resources during the product or system design stage – ecodesign aiming to minimize the environmental impact of a product throughout its life cycle (Kim et al., 2020). To Prieto-Sandoval et al. (2018), IE is seen as a transitional object (notably from the 1960s to the 2000s) serving the shift from a linear to a circular economy, by moving from an explorative or “cowboy” economy to a regenerative one, through restorative and cyclical steps. IE can indeed be helpful to transition towards a more CE, creating different alternatives for the used materials, components, or energy flows, through sharing, reuse, repair, recycling, or energy recovery. Additionally, Saavedra et al. (2018) referred to other IE tools and how they can be used to achieve a sound CE through, e.g., dematerialization, material substitution, pollution prevention, design for environment, or eco-industrial parks. Saavedra et al. (2018) actually studied the theoretical contribution of IE in the transition to CE, indicating several aspects in which IE contributes to the CE, such as conceptual, technical and political aspects. Through their review of 43 publications representing the contributions of IE to the development of CE, the authors identified that the evolution of CE would not be possible without the existence of IE concepts and tools. Interestingly, Blomsma and Brennan (2017) exposed a draft of a state-of-the-art research agenda for IE to contribute to the development of CE. To guide the development of the CE concept towards wide implementation and alignment with sustainable development, they mentioned that further integration of social, economic, and system dynamics theories, with IE is required.

## 4. Results: complementary insights from bibliometrics and text mining

### *4.1. Bibliometric analysis*

Using the “compare” feature of Google Trends (2020), Figure 1 shows the evolution of the global public interest overtime on CE and IE, through web search (worldwide). The values on the Y-axis represent the search interest relative to the highest point on the chart for the given region and time frame. A value of 100 is the peak popularity for the term. A value of 50 means that the term is half as popular. Likewise, a score of 0 means that the term was less than 1% as popular as the peak.

This chart reveals an ever-growing interest in circular economy from 2012 to the present day, while industrial ecology was more popular in the 2000s, 2012 being the turning point when the popularity of CE started to outpace the number of web searches on IE. A similar story can be told in the world of academic publication, as further analyzed hereafter. Actually, this trend also occurred lately – 2016 being the turning point – in the wording used in scientific publications, as illustrated in Figure 2.

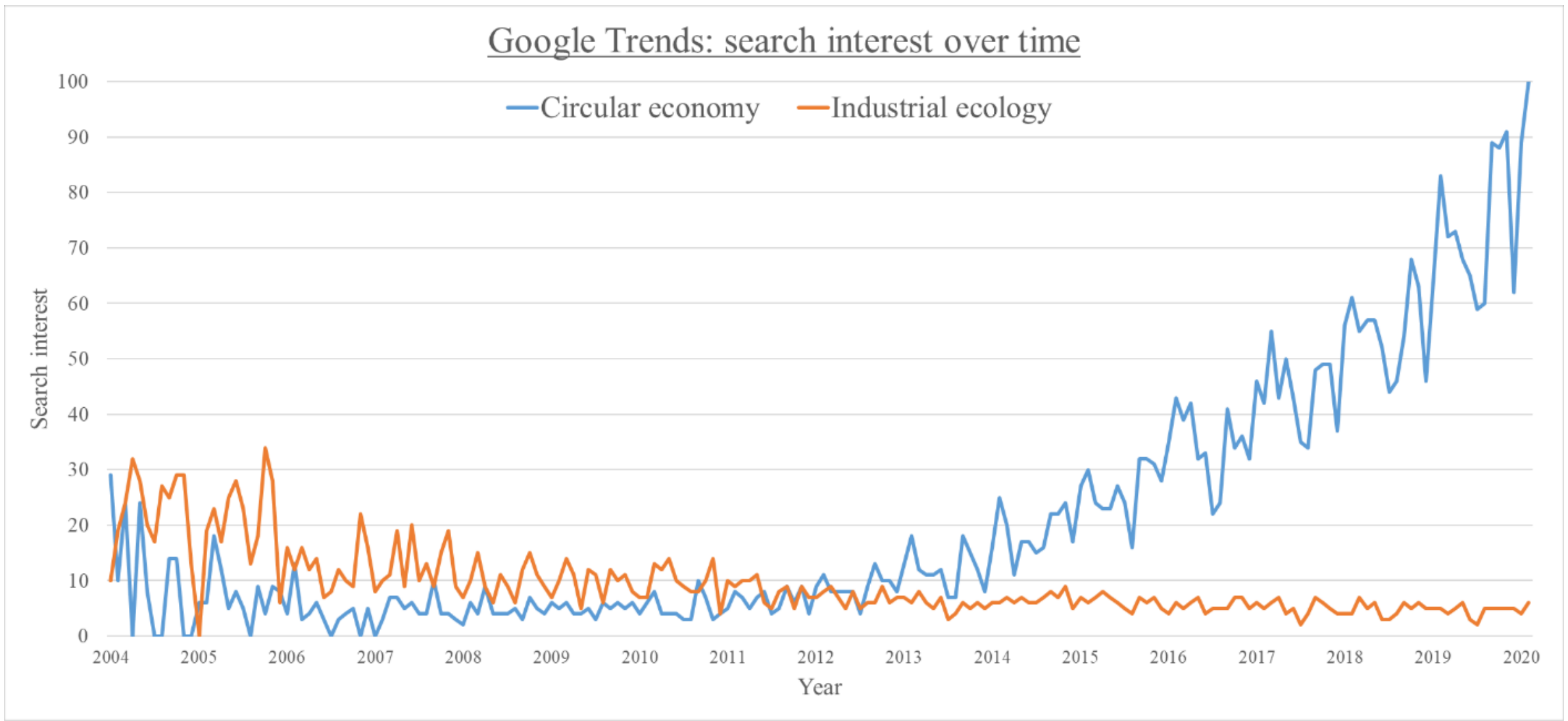

**Figure 1**. Evolution of the search interest on circular economy and industrial ecology (worldwide)

On the one hand, globally, the circular economy concept seems more easily accessible and understandable to the general public (with the straightforward representation of loops, e.g., the butterfly diagram popularized by the Ellen MacArthur Foundation) and the industrial world (driven by economic profit). On the other hand, the industrial ecology concept seems to be more popular within the scientific and research communities, e.g., through the ISIE, that develop methods, tools, and expertise that can serve to the deployment of a circular economy.

Harzing (2007) Publish or Perish software, using Google Scholar queries, has been used to conduct the bibliometric analysis, as illustrated in Figure 2. The six most important international research journals, in terms of the number of publications with CE and/or IE as keywords (Deus et al., 2017), have been considered as a relevant and consistent database for the bibliometric analysis, namely: Journal of Cleaner Production; Journal of Industrial Ecology; Resources, Conservation, and Recycling; Sustainability; Progress in Industrial Ecology, and Waste Management. Here are the key takeaways of this bibliometric analysis: (i) overall, the number of publications with the keyword "industrial ecology" has a steady increased from 2001 to 2019; (ii) while the first consistent publications with the keyword "circular economy" began to emerge in the early 2010s, the number of CE-related publications escalated exponentially since 2016; and, (iii) the fact that the number of publications with the keyword "circular economy" surpasses the ones with the keyword "industrial ecology" does not prevent the latter to still increase.

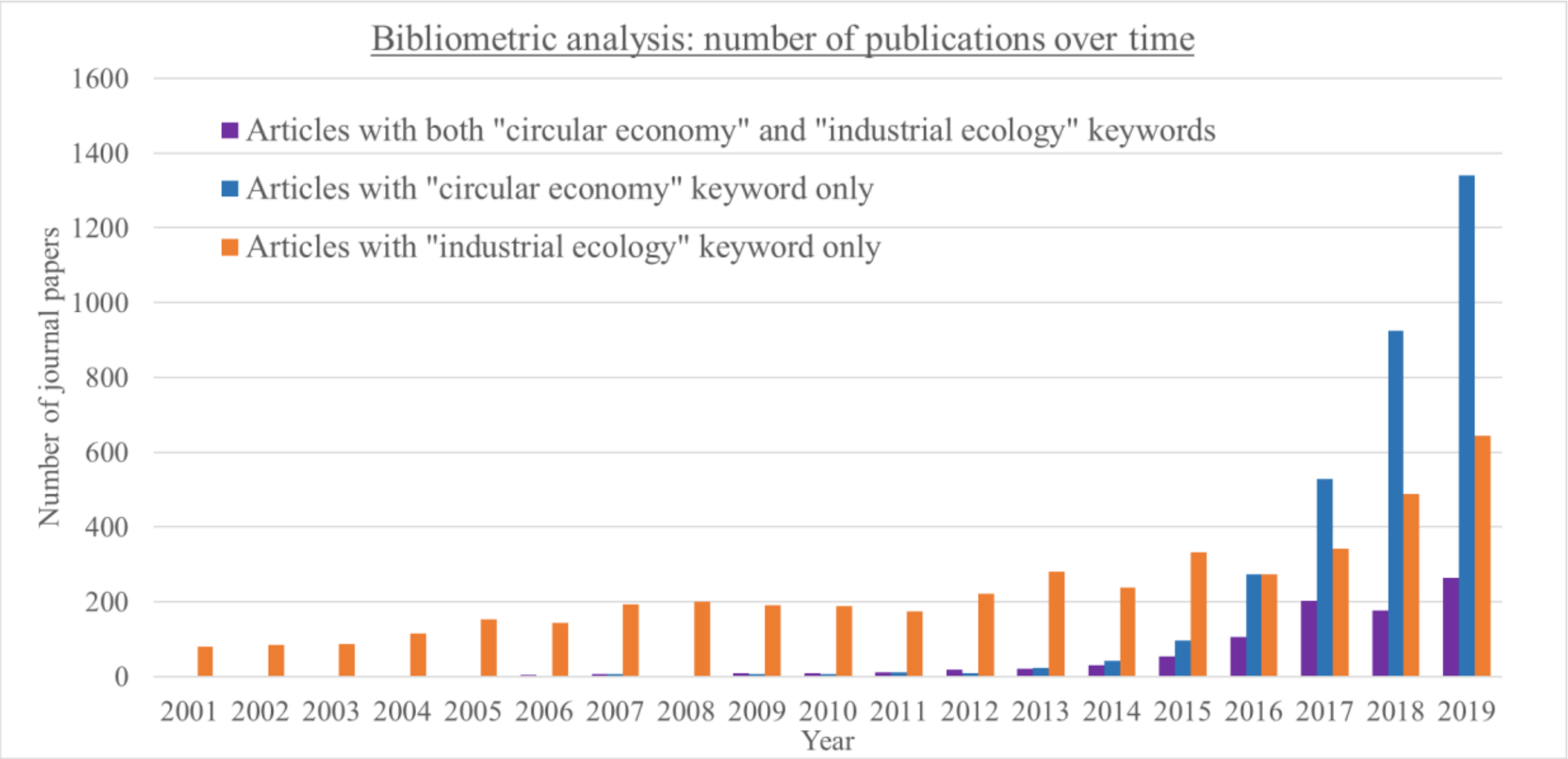

**Figure 2**. Bibliometric analysis of the publications on circular economy and industrial ecology

Furthermore, investigating on the connections between CE, IE and other keywords in journal publications, interesting trends can be deducted. For instance, searching the number of publications (all journals considered, here) having both the keywords "industrial ecology" and "ecodesign", or "circular economy" and "ecodesign", as listed in Table 1, it has been found (e.g., for the year 2016, where the number of publications with the keywords CE and IE are equal, to make a sound comparison), that the ecodesign approach are more often associate with CE strategies that with IE.

Similarly, still using this reference year (2016) for comparison on the same basis, it has been found that: (i) the keywords "business model" and "policy" are more often associated with CE (971, and 179 publications, with these respective keywords, in all journals, reports, and conference proceedings indexed in Google Scholar in 2016) than with IE (510, and 135 publications, respectively); (ii) the keywords “industrial symbiosis” or “eco-park” are more affiliated to IE (750) than to CE (640).

**Table 1**. Number of publications with the keywords circular economy, industrial ecology and/or ecodesign

| Keywords / Year | "industrial ecology", AND "ecodesign" | "circular economy", AND "ecodesign" | "industrial ecology", AND "circular economy", AND "ecodesign" |
|---|---|---|---|
| 2011 | 181 | 21 | 12 |
| 2012 | 222 | 36 | 13 |
| 2013 | 223 | 51 | 19 |
| 2014 | 243 | 104 | 46 |
| 2015 | 244 | 181 | 59 |
| **2016** | **256** | **282** | **92** |
| 2017 | 310 | 427 | 129 |
| 2018 | 297 | 517 | 172 |
| 2019 | 185 | 600 | 167 |

In addition, by analyzing the evolution of CE-related publications produced in the European Union (EU) and in China, Türkeli et al. (2018) identified the most recurrent themes in the joint CE literature between EU and China, including: resource efficiency, indicators, green finance, food and zero waste, governance. Emerging and most important research fields for CE are product life-cycle extensions, new business models, and political economy, while further research is needed on consumer acceptance, rebound effects, social innovation, societal aspects, alternative economies (Türkeli et al., 2018). According to the co-citation network between CE and IE papers, depicted by Saavedra et al. (2018), CE-based research from an IE perspective can further encourage collaborations between these two research communities. Actually, to accomplish a sustainable transition towards a CE at a macro level, a joint work between the business community, policymakers and institutions is fundamental (Saavedra et al., 2018).

*4.2. Text mining analysis*

To complement this state-of-this-art bibliometric analysis, a text mining analysis is conducted on the different definitions of CE, IE, and ecodesign, to further identify and discuss common themes or divergences. Two free online tools are used to perform the text mining: (i) Textalyser, to sort out the most used terms in their definitions, as summarized in Table 2; (ii) Wordle to generate illustrative word clouds based on these definitions, as mapped out in Figure 3. Note that all the 70 definitions of circular economy, 35 definitions of ecodesign, and 13 definitions of industrial ecology used here for the text mining analysis are available in Appendix A of Saidani’s Ph.D. thesis manuscript (2018).

Through this text mining analysis, although definitive interpretations cannot be drawn, some relevant trends are revealed. Regarding the similarities, both CE and IE concepts use substantially the terms “industrial” and “waste” in their definitions. System thinking is also central in both CE and IE definitions. This makes sense because designing out wastes and closing the loop on industrial products needs a holistic understanding and support. Regarding the differences, the noun “resources” is most often used in CE definitions, while the noun “energy” is more employed in IE definitions. A closer focus on “products” is also noticed on the CE definitions. Moreover, the notions of end-of-life strategies appear in CE definitions through the words "recycling" and "reuse", while IE definitions refer more about the production phase with the word "manufacturing". "Design" strategies (e.g., "design for recycling", design for reuse") are also often mentioned in CE definition In this line, ecodesign is interesting as a micro approach aiming to design and develop environmentally sound products, that can support the realization of CE practices in the long run.

Yet, while circular economy and ecodesign shared some common factors (Pole éco-conception, 2016), through the recommendation of closed-loop supply chain, multi-sectorial collaboration and innovative business models, the former is more focus on a continuous economic growth through sustainable consumption, and the latter on reducing environmental footprint. Indeed, ecodesign methods and tools help to design and develop products so that they can be maintained, reused, remanufactured, or recycled through CE loops. Concretely, ecodesign practices could feed the CE paradigm by making products more easily repairable and longer-lasting, e.g., thanks to the selection of proper materials, modules and connections between components.

**Table 2**. Most used terms in the definitions of circular economy, industrial ecology, and ecodesign

| Most cited words | Circular economy | Industrial ecology | Ecodesign |
|---|---|---|---|
| #1 | economy | ecology | design |
| #2 | circular | industrial | environmental |
| #3 | resources | systems | eco |
| #4 | materials | natural | product |
| #5 | economic | energy | life |
| #6 | waste | materials | cycle |
| #7 | system | economic | process |
| #8 | use | waste | impacts |
| #9 | products | human | development |
| #10 | value | environmental | products |
| #... | energy, production, recycling, development, consumption, industrial, reuse, design | products, ecosystems, manufacturing, industry | environment, integration, stages, reduce, account |

Note that as a first analysis of these definitions, for simple and straightforward interpretations, only the term frequency mining technique was used, i.e., extracting how frequently a word appears in a document, and its relative importance to the whole set of words. One key limitation is that it can disregard or overlook a lot of information on the syntax and words that are semantically similar. Further studies can be conducted using a more complex text mining technique, such as a cluster analysis (Talib et al., 2016), e.g. to analyze more specifically the associations between keywords.

## 5. Discussion and perspectives

Through bibliometric and text mining analyses, this piece of research brought new insights on the positioning of circular economy within the industrial ecology field, and *vice versa*. Understanding the evolution and recent trends, as well as clarifying the similarities and divergences between these concepts is essential to ensure proper communication, and adequate implementation, for researchers, industrialists, or policymakers, working on circular economy and industrial ecology. Combining these new findings with the corpus of literature studying these relations, further discussions can be made on the appropriate representations and usages of CE and IE concepts.

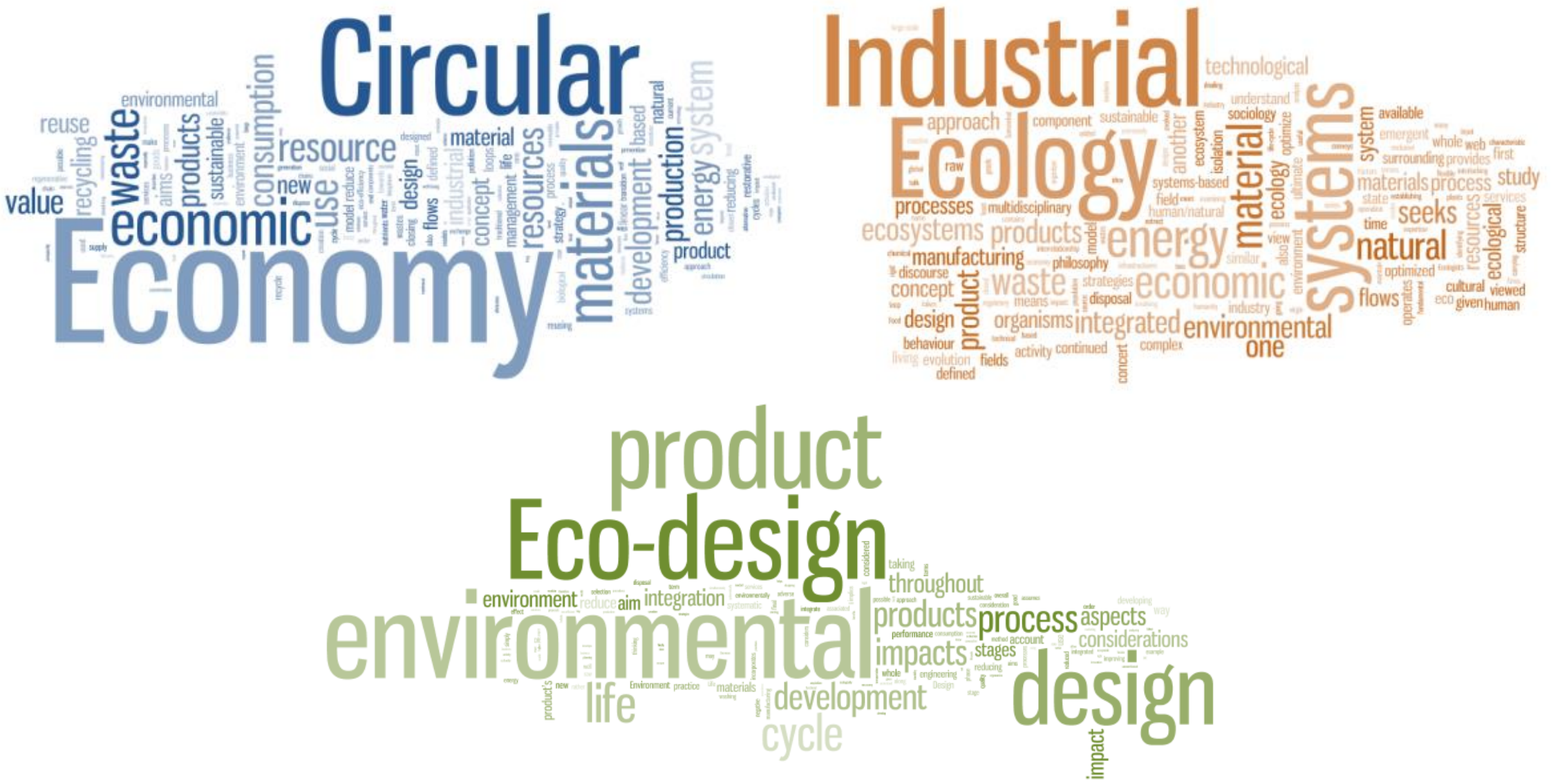

**Figure 3**. Word clouds of circular economy, industrial ecology and ecodesign definitions

What is novel and exciting for the CE relies on the traction it gained relatively rapidly – compared to the IE, which barely had a couple of mentions in industrial press releases back in the 1990s or 2000s (Bocken et al., 2017) – among businesses, industrial practitioners, and policy-making communities (Blomsma and Brennan, 2017; Korhonen et al., 2018a). Yet, even if an increasing number of firms (e.g., Caterpillar, Renault, or Unilever) are setting and communicating on CE strategies and associated targets implementing in their operational practices, there is still a long way to go for widespread adoption of CE- or IE-centered mindset in many businesses or industrial sectors (Bocken et al., 2017; Masi et al., 2017). In fact, although lot of ecodesign tools are available, their implementation in practice still requires some efforts (Mathieux et al., 2020).

As such, today is a timely period for researchers, and for sustainability or corporate social responsibility departments in companies, to further build on this momentum and foster the use of ecodesign, industrial ecology tools, to accelerate the transition towards a circular and sustainable economy. Actually, in addition to the boost that CE can benefit from the its inclusion in more and more political agendas (Geissdoerfer et al., 2017), the previously developed – and now available in various formats – tools and methods for industrial ecology (e.g., life cycle assessment, material flow analysis), for innovative business models (e.g., triple-layer business model canvas), are valuable resources that can serve to operationalize the CE

In all, these concepts – either circular economy, industrial ecology, ecodesign – should work hand in hand and feed each other to enable and catalyze the sound sustainable management of natural resources in the 21st century socio-economic world. For instance, the recent integration of CE requirements helped to move forward the EU EcoDesign Directive (Peiró et al., 2020), which was mostly focused on regulating energy consumption, with a large untapped potential into material efficiency (Mathieux et al., 2020). As both CE and IE are increasingly acknowledged to be suitable instruments for sustainable development (Geissdoerfer et al. 2017; Korhonen et al., 2018b; Schroeder et al., 2019), it would be relevant in future work to assess more quantitatively how and to what extent CE incentives, strategies, and IE projects are contributing to sustainable development, e.g., through the appropriate combination of circularity and sustainability indicators (Saidani et al., 2019; Kravchenko et al., 2020), in order to foster and monitor such actions.

**Acknowledgements**

Some sections of this paper are partially based on the introduction section of the Ph.D. thesis manuscript of the lead author (Saidani, 2018). These parts have been reworked, updated, and augmented to provide an independent and self-contained piece of research providing new and stand-alone insights for the research, industrial, and policy-making communities, working on circular economy implementation.

**References**

ADEME. (2014). Methodological Guide to the Development of Regional Circular Economy Strategies in France. French Environment & Energy Management Agency.

Blomsma, F. (2018). Collective 'action recipes' in a circular economy – On waste and resource management frameworks and their role in collective change, *Journal of Cleaner Production*, 199, 969–982.

Blomsma, F., Brennan, G. (2017). The Emergence of Circular Economy: A New Framing Around Prolonging Resource Productivity. *Journal of Industrial Ecology*, 21, 603–614.

Bocken, N.M.P., Ritala, P., Huotari, P. (2017). The Circular Economy: Exploring the Introduction of the Concept Among S&P 500 Firms. *Journal of Industrial Ecology*, 21, 487–490.

Bonet, D., Petit, I., Lancini, A. (2014). L'économie circulaire : quelles mesures de la performance économique, environnementale et sociale ? *Revue française de gestion industrielle*, 33, 4.

Bourg, D., Erkman, S. (2003). Perspectives on Industrial Ecology. Greenleaf Publishing, Sheffield.

Buclet, N. (2015). Ecologie industrielle et économie circulaire: définitions et principes. *Management & Société*, 27–41.

Cecchin, A., Salomone, R., Deutz, P., Raggi, A., Cutaia, L. (2020). Relating Industrial Symbiosis and Circular Economy to the Sustainable Development Debate. In: Industrial Symbiosis for the Circular Economy, Springer, Cham.

Deus, R.M., et al. (2017). Trends in publications on the circular economy. *Revista Espacios*, 38, 58.

Ehrenfeld, J. (2004). Industrial ecology: a new field or only a metaphor? *Journal of Cleaner Production*, 12, 825–831.

Ellen MacArthur Foundation. (2015). www.ellenmacarthurfoundation.org/circular-economy/concept/schools-of-thought

Erkman, S. (1997). Industrial ecology: an historical view. *Journal of Cleaner Production*, 5(1–2), 1–10.

Figuière, C., Chebbi, A. (2016) Écologie Industrielle (EI) et Économie Circulaire (EC). Concurrentes ou complémentaires ? XXXIIèmes journées du développement ATM 2016, Lille, France.

Frosch, R.A., Gallopoulos, N.E. (1989). Strategies for Manufacturing. *Scientific American*, 261, 144–152.

Gaikwad, S.V., Chaugule, A., Patil, P. (2014). Text mining methods and techniques. *International Journal of Computer Applications*, 85, 17.

Garner, A., Keoleian, G.A. (1995). Industrial Ecology: An Introduction. Pollution Prevention and Industrial Ecology. National Pollution Prevention Center for Higher Education, University of Michigan.

Geissdoerfer, M., Savaget, P., Bocken, N.M., Hultink, E.J. (2017). The Circular Economy—A new sustainability paradigm? *Journal of Cleaner Production*, 143, 757–768.

Google Trends. (2020). www.trends.google.com/trends/explore?date=all&q=circular%20economy,industrial%20ecology

Harzing, A.W. (2007). Publish or Perish. www.harzing.com/resources/publish-or-perish

Hirsch, P.M., Levin, D.Z. (1999). Umbrella advocates versus validity police: A life-cycle model. *Organization Science*, 10 (2), 199–212.

International Society for Industrial Ecology. (2015). A Short History of Industrial Ecology. www.is4ie.org/about/history

Jelinski, L.W., Graedel, T.E., Laudise, R.A., McCall, D.W., Patel, C.K. (1992). Industrial ecology: concepts and approaches. *Proceedings of the National Academy of Sciences*, 89(3), 793–797.

Kim, H., Cluzel, F., Leroy, Y., Yannou, B., Yannou-Le Bris, G. (2020). Research perspectives in ecodesign. *Design Science*, 6.

Kirchherr, J., Reike, D., Hekkert, M. (2017). Conceptualizing the circular economy: An analysis of 114 definitions. *Resources, Conservation and Recycling*, 127, 221–232.

Korhonen, J., Honkasalo, A., Seppälä, J. (2018a). Circular economy: the concept and its limitations. *Ecological economics*, 143, 37–46.

Korhonen, J., Nuur, C., Feldmann, A., Birkie, S.E. (2018b). Circular economy as an essentially contested concept. *Journal of Cleaner Production*, 175, 544–552.

Kovács, G. (2017). Circular economy vs. closed loop supply chains: what is new under the sun? Constructing A Green Circular Society, First edition, Nov. 2017.

Kravchenko, M., Pigosso, D.C., McAloone, T.C. (2020). A Procedure to Support Systematic Selection of Leading Indicators for Sustainability Performance Measurement of Circular Economy Initiatives. *Sustainability*, 12(3), 951.
Leydesdorff, L., Van den Besselaar, P. (1997). Scientometrics and communication theory: Towards theoretically informed indicators. *Scientometrics*, 38 (1), 155–174.
Lifset, R., Graedel, T.E. (2002). Industrial ecology: goals and definitions. A handbook of industrial ecology, 3–15.
Masi, D., Day, S., Godsell, J. (2017). Supply chain configurations in the circular economy: A systematic literature review. *Sustainability*, 9(9), 1602.
Mathieux, F., et al. (2020). Ten years of scientific support for integrating CE requirements in the EU EcoDesign Directive: overview and lessons learnt. *Procedia CIRP*, In press.
Peiró, L.T., Polverini, D., Ardente, F., Mathieux, F. (2020). Advances towards circular economy policies in the EU: The new Ecodesign regulation of enterprise servers. *Resources, Conservation and Recycling*, 154, 104426.
Pigosso, D.C.A, Zanette, E.T., Filho, A.G., Ometto, A.R., Rozenfeld, H. (2010). Ecodesign methods focused on remanufacturing, *Journal of Cleaner Production*. 18(1), 21–31.
Pole éco-conception. (2016). www.eco-conception.fr/static/economie-circulaire.html
Prieto-Sandoval, V., Jaca, C., Ormazabal, M. (2018). Towards a consensus on the circular economy, *Journal of Cleaner Production*, 179, 605–615.
ResearchGate. (2019). https://www.researchgate.net/post/Are_Circular_Economy_and_Industrial_Ecology_the_same_concept_And_if_not_which_are_the_differences
Røine, K. (2000). Does industrial ecology provide any new perspectives? Reports and Communications from Norwegian University of Science and Technology, Industrial Ecology Programme, Report n° 3/2000.
Saavedra, Y.M.B., Iritani, D.R., Pavan, A.L.R., Ometto, A.R. (2018). Theoretical contribution of industrial ecology to circular economy, *Journal of Cleaner Production*, 170, 1514–1522.
Sacchi Homrich, A., Galvão, G., Gamboa Abadia, L., Carvalho, M.M. (2018). The circular economy umbrella: Trends and gaps on integrating pathways. *Journal of Cleaner Production*, 175, 525–543.
Saidani, M. (2018), Monitoring and advancing the circular economy transition – Circularity indicators and tools applied to the heavy vehicle industry. PhD Thesis, Université Paris-Saclay.
Saidani, M., Yannou, B., Leroy, Y., Cluzel, F., Kendall, A. (2019). A taxonomy of circular economy indicators. *Journal of Cleaner Production*, 207, 542–559.
Schroeder, P., Anggraeni, K., Weber, U. (2019). The relevance of circular economy practices to the sustainable development goals. *Journal of Industrial Ecology*, 23(1), 77–95.
Talib, R., Hanif, M.K., Ayesha, S., Fatima, F. (2016). Text mining: techniques, applications and issues. *International Journal of Advanced Computer Science and Applications*, 7(11), 414–418.
Türkeli, S., et al. (2018). Circular economy scientific knowledge in the European Union and China: A bibliometric, network and survey analysis (2006–2016). *Journal of Cleaner Production*, 197, 1244–1261.